\documentclass[aps,pra,twocolumn,nopacs,floatfix]{revtex4-2}
\usepackage{epsfig}
\usepackage{graphicx}
\usepackage{dcolumn}
\usepackage{amsthm,amsmath,amsfonts}
\usepackage{mathtools}
\usepackage{comment}
\usepackage[colorlinks]{hyperref}
\usepackage{xcolor}
\usepackage{orcidlink}
\usepackage[normalem]{ulem}

\begin{document}

\title{Demonstrating Correlation Trends in the Electric Dipole Polarizabilities of Many Low-lying States in Cesium (Cs I) through First-principle Calculations}

\author{A. Chakraborty \orcidlink{0000-0001-6255-4584}}
\email{arupc794@gmail.com}

\author{B. K. Sahoo \orcidlink{0000-0003-4397-7965}}
\email{bijaya@prl.res.in}

\affiliation{Atomic, Molecular and Optical Physics Division, Physical Research Laboratory, Navrangpura, Ahmedabad 380009, India}

\begin{abstract}
Electron correlation and higher-order relativistic effects are probed in the evaluation of scalar and tensor static electric dipole (E1) polarizabilities ($\alpha_d$) of several even- and odd-parity states in cesium (Cs) using the Dirac-Hartree-Fock (DHF) method, second-order perturbation theory (MBPT(2)), third-order perturbation theory (MBPT(3)), random phase approximation (RPA), and singles and doubles approximated relativistic coupled-cluster (RCCSD) method. To account for perturbation due to odd-parity E1 operator on the atomic orbitals, calculations are carried out in the linear response approach. Our final $\alpha_d$ values, with the estimated uncertainties, show reasonably good agreement with the previous calculations and available experimental results. Differences among the DHF, MBPT(2), MBPT(3) and RPA results indicate pair-correlation (PC) effects play major roles than the core-polarization (CP) effects in the determination of $\alpha_d$ values in Cs. From the differences among the MBPT(3) and RCC results, we find correlations among the PC and CP effects and double CP effects together are also significant in these calculations. Contributions from the Breit interactions are found to be quite large in the high-lying states.    
\end{abstract}

\date{\today}

\maketitle 

\section{Introduction}\label{secin}

Precise estimations of electric dipole (E1) polarizabilities ($\alpha_d$) are essential to estimate systematic effects due to light shifts in the high-precision measurements using atomic systems \cite{Palchikov2003, Yudin2011}. Particularly, their accurate values are useful in the atom trapping, atomic clocks, quantum computing and testing fundamental physics experiments \cite{Schlosser2001, Williams1997, DiVincenzo2000, Negretti2011, Kozlov2018}. Alkali atoms are usually preferred to be undertaken to conduct various studies involving high-precision measurements owing to their simple electronic structures and well-characterized properties \cite{Liew2004, Saffman2010}. Particularly, the cesium (Cs) atom is preferred in the experiments which is the heaviest non-radioactive alkali atom. $^{133}$Cs based atomic clocks help defining the SI second and support metrology, including space missions where precise timekeeping is crucial \cite{Lammerzahl2004, Taylor2001, Carr2016}. This is also the only atomic system in which parity violation (PV) amplitude has been measured within 0.5\% accuracy \cite{Wood1997} and help constraining physics beyond the Standard Model of particle physics and extracting nuclear anapole moment \cite{Bouchiat1997, Ginges2004, Chakraborty2024}.

Given its importance in both applied and fundamental research, Cs has been the subject of extensive theoretical and experimental studies. Over the past five decades, numerous investigations have been conducted to determine the $\alpha_d$ values of both the ground and excited states of Cs \cite{Safronova2004, Safronova1999, Chakraborty2023, Iskrenova2007, Safronova2016, Tang2014, Bouchiat1983, Amini2003, Bennett1999, Quirk2024, Weaver2012, Gunawardena2007, Antypas2011, Hunter1988, Hunter1992, Toh2014, Domelunksen1983, Tanner1988}. Despite the vast body of work, persistent discrepancies exist among various theoretical predictions and experimental results, leading to unresolved controversies in the field \cite{Safronova2004, Bennett1999, Quirk2024, Hunter1988, Hunter1992, Tanner1988}. One of the most enduring and debated issues is related to the polarizabilities of the $6P_{1/2}$ and $6P_{3/2}$ states \cite{Safronova2004, Sahoo2016}. Polarizabilities of these states show significant inconsistencies among the theoretical and experimental values and among different measurements carried out independently \cite{Safronova2004, Hunter1988, Hunter1992, Tanner1988}. Notably, two independent measurements of polarizabilities of the $6P_{1/2}$ and $6P_{3/2}$ states report values differing by approximately 3\% and 1.5\%, respectively \cite{Hunter1988, Hunter1992, Tanner1988}. A similar situation arises in the case of the $7S$ state, where experimental findings have introduced further debates on these studies. In a recent study, Quirk {\it et al.} \cite{Quirk2024} reported a polarizability value of $6207.9$ atomic units (a.u.) for the $7S$ state. However, this result deviates significantly from previously established theoretical predictions and an earlier experimental value reported to be approximately 6238 a.u. \cite{Safronova2016, Tang2014, Bennett1999}. This level of disagreement underscores the need for employing more accurate theories or carrying out further high-precision measurements of $\alpha_d$ in the low-lying states to reconcile these inconsistencies and refine our understanding of Cs atomic structure.

The earlier calculations were carried out by employing basically linearized version of a relativistic coupled-cluster (RCC) method \cite{Safronova1999, Safronova2016, Iskrenova2007} or Dirac-Hartree-Fock (DHF) method with core-polarization (CP) potential approach \cite{Tang2014}. The high-precision $\alpha_d$ values reported from these calculations are mostly obtained using the sum-over-states approach. In this approach, the dominant valence correlation contributions are evaluated by combining the high-precision E1 matrix elements from either calculations or measurements with the experimental energies. The remaining contributions from the valence correlations due to E1 matrix elements involving high-lying states are estimated using either mean-field calculations at the DHF method or lower-order many-body perturbation theory (MBPT). Moreover, contributions from the occupied orbitals, that cannot be estimated accurately in the sum-over-states approach, were considered through {\it ab initio} calculations using MBPT or random phase approximation (RPA). Such mixed approaches to determine $\alpha_d$ values have many shortcomings from the interest of theoretical studies -- (i) this approach does not treat all the correlation contributions on equal footing, (ii) it cannot take into account correlations among contributions that are evaluated separately, (iii) propagation of correlation effects from lower- to higher-order methods cannot be probed in different states explicitly, and (iv) it cannot account for contributions from double-core-polarization (DCP) effects that involve intermediate states with double excitations. This calls for employing first-principle approaches to calculate the $\alpha_d$ values systematically; particularly in Cs in view of the aforementioned debates on some of its reported results. 

Among all the many-body methods commonly employed to carry out atomic calculations, the RCC theory is considered to be more powerful. This method not only takes into account electron correlations more rigorously it also obeys size-consistent behavior. However, this method cannot be directly employed to determine $\alpha_d$ values of atomic systems in the spherical coordinate system. Thus, we considered linear response (LR) approach in the RCC theory framework in which atomic orbitals are perturbed explicitly due to the odd-parity E1 operator as discussed in our previous works \cite{Sahoo2007, Sahoo2009, Chakraborty2022, Chakraborty2025}. In this approach, core, core-valence and valence correlation effects are treated on equal footing. It also accounts CP, pair-correlation (PC), DCP and their correlations to all-orders. In order to understand propagation of correlation effects in the evaluation of $\alpha_d$ values, we present results at the DHF, second-order MBPT (MBPT(2)) method, third-order MBPT (MBPT(3)) method, RPA, and singles and doubles approximated RCC (RCCSD) method. In the LR approach, the MBPT(2) method contains the lowest-order CP contributions while RPA includes CP contributions to all-orders. Similarly, the MBPT(3) method contains the lowest-order PC effects and the RCCSD method contains all-order PC effects. Moreover, the RCCSD method contains correlations among the CP and PC effects as well as DCP contributions. Therefore, it is possible to understand importance of CP, PC and DCP effects by analyzing trends of $\alpha_d$ values in different states of Cs through the above mentioned methods.    

The paper is organized as follows: Sec. \ref{secth} introduces the scalar and tensor components of $\alpha_d$ and Sec. \ref{secme} outlines many-body methods employed in the LR approach to compute $\alpha_d$ of different states of Cs. The next section presents and discusses results by comparing them with the earlier studies and highlight roles of important correlation effects in the accurate evaluation of $\alpha_d$ following the conclusion of the work. Unless specified otherwise, all results are reported in a.u..

\section{Theory} \label{secth}

For an isolated atom, parity is a good quantum number. However, when an atom is placed in an external electric field, it loses its spherical symmetry leading to shifts in its energy levels. For an atomic state $|J_nM_n\rangle$, where $n$ is the principal quantum number, the dominant energy shift due to a static electric field $\vec {\mathbb{E}} = {\cal E}_0 \hat{\epsilon}$ arises from the second-order effect, given by
\begin{eqnarray}
\Delta E^{(2)}(J_n,M_n) = -\frac{1}{2} \alpha_d(J_n,M_n) {\cal E}_0^2,
\end{eqnarray}
where $\alpha_d(J_n,M_n)$ is the static electric dipole polarizability of the corresponding atomic state. As can be seen from the above equation, $\alpha_d(J_n,M_n)$ depends on $J_n$ and $M_n$. Using the $M_n$-dependent factors, it yields \cite{Manakov1986, Stalnaker2006}
\begin{eqnarray}
\alpha_d(J_n,M_n)=\alpha_d^{S}(J_n) + \frac{3M_n^2-J_n(J_n+1)}{J_n(2J_n-1)}\alpha_d^{T}(J_n). 
\end{eqnarray}
Here $\alpha_d^{S}(J_n)$ and $\alpha_d^{T}(J_n)$ are called as the scalar and tensor polarizabilities, respectively. These $M_n$ independent quantities can be written in terms of reduced matrices using Spherical tensors of angular momentum operators as \cite{Manakov1986}
\begin{eqnarray}
 \alpha_d^S(J_n) &=& C_0\sum_{k}\frac{|\langle J_n||{\bf D} ||J_k \rangle|^2}{E^{(0)}_{J_n} - E^{(0)}_{J_k}}
 \label{eqas}
\end{eqnarray}
and
\begin{eqnarray}
 \alpha_d^T(J_n)&=&\sqrt{\frac{40J_n(2J_n-1)}{3(J_n+1)(2J_n+3)(2J_n+1)}}\sum_k (-1)^{J_n+J_k+1}\nonumber\\
 &&\times\left\{ \begin{array}{ccc}
                J_n& 2 & J_n\\
                1 & J_k &1 
 \end{array}\right\} (-1)^{J_n-J_k} \frac{|\langle J_n|| {\bf D} ||J_k \rangle|^2}{E^{(0)}_{J_n} - E^{(0)}_{J_k}} \nonumber\\
 &=& \sum_{k} C_k \frac{|\langle J_n|| {\bf D} ||J_k \rangle|^2}{E^{(0)}_{J_n} - E^{(0)}_{J_k}},  
 \label{eqat}
\end{eqnarray}
where $C_0 =  - \frac{2}{3(2J_n+1)}$, $C_k = \sqrt{\frac{40J_n(2J_n-1)}{3(J_n+1)(2J_n+3)(2J_n+1)}}  \times  (-1)^{J_n+J_k+1} \left\{ \begin{array}{ccc}
                    J_n& 2 & J_n\\
                  1 & J_k &1 
 \end{array}\right\} $
with the curly bracket symbol denoting the 6j coefficient and  ${\bf D}=\sum_{q=-1}^{1} d^{(1)}_q$ is the E1 operator. From the angular momentum selection rules, it can be shown that $ \alpha_d^T(J_n)$ will be non-zero only for the states $J_n>1/2$. 

\begin{figure}[t!]
\centering
\includegraphics[width=85mm,height=75mm]{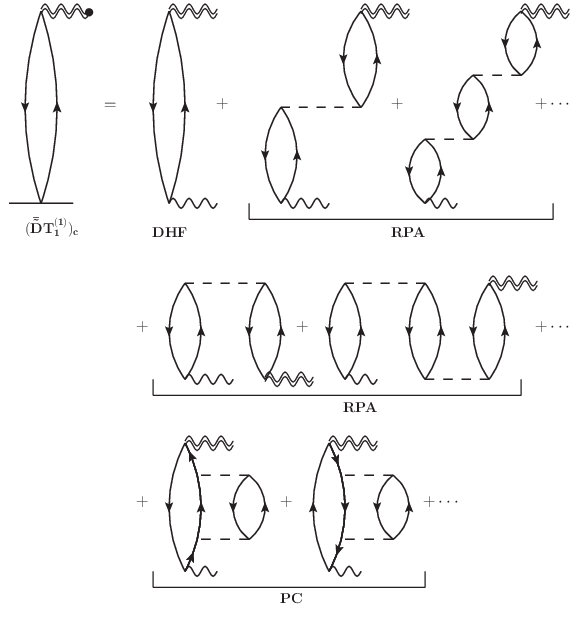}
\caption{Breakdown of the RCC term $(\bar{\tilde{D}}T_1^{(1)})_c$ to the MBPT terms through Goldstone diagram representation. Here, single arrows going down and up mean occupied and virtual orbitals, respectively. The dotted line corresponds to the residual $V_{res}$ interaction from the atomic Hamiltonian, the single curly line represents the E1 operator $D$, the double curly line corresponds to the effective E1 operator $\tilde{D}$, and the double curly line with a bullet denotes $\bar{\tilde{D}}$ that are shown using the MBPT diagrams.}
\label{cor1}
\end{figure}

By dividing electrons into core and valence electrons, correlation contributions to $\alpha_d^{S/T}$ can be divided into \cite{Arora2012, Kaur2015} 
\begin{eqnarray}
\alpha_d^{S/T}=\alpha_{c}^{S/T}+\alpha_{cv}^{S/T}+\alpha_{v}^{S/T},
\end{eqnarray}
where terms with subscripts $c$, $cv$, and $v$ denote contributions from core, core-valence, and valence correlation, respectively. Due to appearance of the 6j coefficient in $C_k$, the core contribution to $\alpha_d^T$ will be zero, as for the core sector, $J=0$. 

\section{Methodology} \label{secme}

It is relatively easier to evaluate the $\alpha_{v}^{S}$ and $\alpha_{v}^{T}$ values of atomic states in the sum-over-states approach using formulas given by Eqs. (\ref{eqas}) and (\ref{eqat}) with the knowledge of reduced matrix elements $\langle J_n||D||J_k \rangle$ and excitation energies of dominantly contributing low-lying transitions. Since $\langle J_n||D||J_k \rangle$ amplitudes and their excitation energies involving occupied or continuum orbitals cannot be evaluated explicitly using many-body methods, a sum-over-states approach cannot be used to determine $\alpha_c^{S/T}$,  $\alpha_{cv}^{S/T}$ and continuum contributions to $\alpha_{v}^{S/T}$. Thus, they are estimated often using lower-level methods like DHF method or RPA; particularly in Cs. For comprehensive and accurate estimations of these contributions to $\alpha_d^{S/T}$, we express their expressions in the LR approach as \cite{Chakraborty2025} 
\begin{eqnarray}
 \alpha_d^{S/T}&=&\langle \Psi_n^{(0)}|\tilde{D}^{S/T}|\Psi_n^{(1)} \rangle+\langle \Psi_n^{(1)}|\tilde{D}^{S/T}|\Psi_n^{(0)} \rangle\nonumber\\
 &=&2\langle \Psi_n^{(0)}|\tilde{D}^{S/T}|\Psi_n^{(1)} \rangle ,
\end{eqnarray}
where $|\Psi_n^{(0)} \rangle$ and $|\Psi_n^{(1)} \rangle$ denote the unperturbed and first-order perturbed wave functions of the system, respectively. These wave functions are solutions to the atomic Hamiltonian ($H_{at}$) and the perturbative corrections due to the E1 interaction. The term $\tilde{D}^{S/T}$ refers to the effective dipole operators for scalar and tensor components, given by $\tilde{D}^S = C_0 D$ and $\tilde{D}^T = \sum_k C_k D$.

The first-order perturbed wave function $|\Psi_n^{(1)} \rangle$ can be obtained by solving the following inhomogeneous equation
\begin{eqnarray}
 (H_{at}-E_n^{(0)})|\Psi_n^{(1)}\rangle=-D|\Psi_n^{(0)}\rangle.
\end{eqnarray}
Solution of this equation using a given method would ensure that all the correlation contributions to $\alpha_d^{S/T}$ are treated equally. However, accurate determination of these contributions depend on the choice of atomic Hamiltonian and many-body method. To demonstrate dependency of results with choice of a method, we present results from the DHF, MBPT(2), MBPT(3), RPA and RCCSD methods. These methods are discussed briefly below, while details of these methods can be found elsewhere \cite{Sahoo2007, Singh2014, Sahoo2017, Chakraborty2023-2, Chakraborty2025}. 

To begin with, we consider $H_{at}$ at the Dirac-Coulomb approximation, given by (in a.u.)
\begin{eqnarray}\label{eq:DC}
H_{at} &\equiv& \sum_i \left [c {\vec \alpha}_i^D \cdot {\vec p}_i+(\beta_i^D-1)c^2+V_n(r_i)\right ] +\sum_{i,j>i}\frac{1}{r_{ij}}. \nonumber \\
\end{eqnarray}
Here, $\alpha^D$ and $\beta^D$ are the Dirac matrices, $\vec{p}$ is the single-particle momentum operator, $V_n(r)$ represents the nuclear potential felt by an electron, and $\frac{1}{r_{ij}}$ represents the Coulomb repulsion between two electrons. We also estimate corrections due to the Breit interaction and lowest-order Quantum Electrodynamics (QED) effects. The QED effects include the lowest-order vacuum polarization effect, described through the Uehling potential and Wichmann-Kroll potential, as well as the self-energy effect described by the magnetic and electric form factors \cite{Flambaum2005, Sahoo2016}.

\begin{figure}[t!]
\centering
\includegraphics[width=85mm,height=75mm]{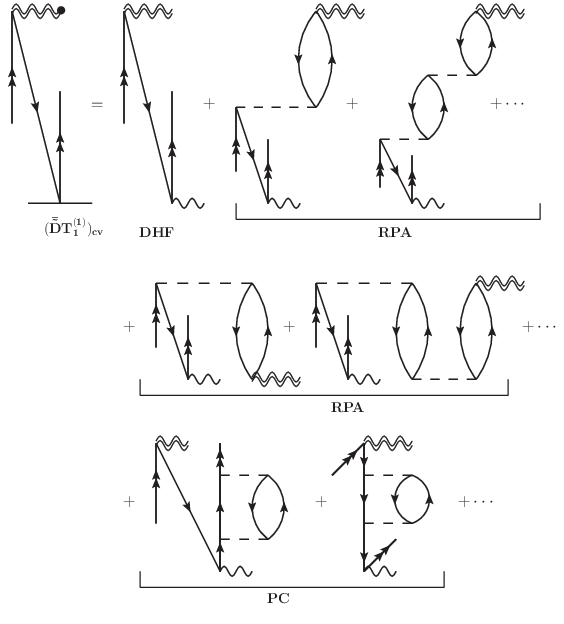}
\caption{Breakdown of the Goldstone diagram corresponding to $(\bar{\tilde{D}}T_1^{(1)})_{cv}$ in terms of MBPT diagrams. Here, the double arrow represents the valence orbital while other symbols are same as described in Fig. \ref{cor1}.}
\label{corv}
\end{figure}

Since all the considered atomic states of Cs have the closed-core $[5p^6]$ and a valence orbital with different parity and angular momentum, we consider the $V^{N_e-1}$ potential formalism, with $N_e=55$ being the number of electrons, to produce the initial wave function, $|\Phi_0 \rangle$, using the DHF Hamiltonian, $H_{DHF}$. This choice of DHF wave function helps to produce wave functions of all the considered states in the Fock-space approach by appending corresponding valence orbital as described below. However, this choice of $V^{N_e-1}$ misses out the correlations among the core electrons and the valence electron. These contributions are later accounted through the core-valence contributions. Again, atomic orbitals generated in this work respect parity as a good quantum number because of which electric dipole interaction cannot be included self-consistently at the DHF level. Instead, we adopt a linear response approach to determine the $\alpha_d$ values by perturbing the wave functions due to $D$. In this approach, the missing orbital relaxation effects that represent the core-polarization interactions to all-orders at the DHF level appear through the RPA-like diagrams in the linear response approach. Thus, the differences in the contributions from the DHF and RPA values to $\alpha_d$ would be equivalent to the orbital relaxation effects in the present work. 

To connect results given at different levels of approximations in the method, we express the final (unperturbed) wave function of the closed-core, $|\Psi^{(0)}_0\rangle$, due to $H_{at}$ by
\begin{eqnarray}
|\Psi_0^{(0)}\rangle=\Omega_0^{(0)}|\Phi_0\rangle ,
\end{eqnarray}
where $\Omega_0^{(0)}$ is referred to as the wave operator. Obviously, $\Omega_0^{(0)}=1$ in the DHF method while it accounts for the electron correlation effects arising from the residual interaction $V_{res}=H_{at}-H_{DHF}$ in a many-body method. 

In order to obtain the desired wave function of an intended state of Cs, we append the required valence orbital, $v$, to the closed-core configuration in the next step by defining the modified DHF wave function as $|\Phi_v \rangle = a_v^{\dagger} |\Phi_0 \rangle$. This follows the final unperturbed wave function in the wave operator formalism 
\begin{eqnarray} 
|\Psi_v^{(0)}\rangle=(\Omega_0^{(0)}+\Omega_v^{(0)})|\Phi_v\rangle ,
\end{eqnarray} 
where $\Omega_v^{(0)}$ is responsible for accounting the correlation effects involving the electron from the valence orbital $v$. Similarly, the corresponding first-order perturbed wave functions can be expressed as
\begin{eqnarray}
|\Psi_0^{(1)}\rangle=\Omega_0^{(1)}|\Phi_0\rangle 
\end{eqnarray}
and
\begin{eqnarray} 
|\Psi_v^{(0)}\rangle=(\Omega_0^{(1)}+\Omega_v^{(1)})|\Phi_v\rangle ,
\end{eqnarray} 
where superscript (1) on wave operators stands for the first-order perturbation. 

\begin{figure}[t!]
\centering
\includegraphics[width=85mm,height=100mm]{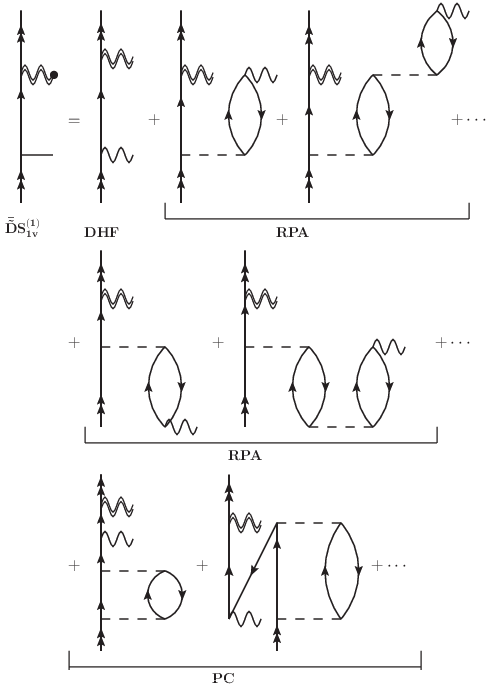}
\caption{Breakdown of the RCC term $\bar{\tilde{D}}S_{1v}^{(1)}$ to the MBPT terms through Goldstone diagrams. All the symbols used here represent the same operators as those in the previous figures.}
\label{s1v}
\end{figure}

In the DHF method, like the $\Omega_0^{(0)}$ operator $\Omega_v^{(0)}=1$ but 
\begin{eqnarray}
\Omega_0^{(1)} = \sum_{ap} \frac{ \langle \Phi_a^p | D | \Phi_0 \rangle} { \epsilon_p - \epsilon_a} a_p^{\dagger} a_a = \sum_{ap,k} \frac{ \langle \phi_p | d^{(1)}_k | \phi_a \rangle} { \epsilon_p - \epsilon_a} a_p^{\dagger} a_a 
\end{eqnarray}
and 
\begin{eqnarray}
\Omega_v^{(1)} = \sum_p \frac{ \langle \Phi_v^p | D | \Phi_v \rangle} { \epsilon_p - \epsilon_v} a_p^{\dagger} a_v = \sum_{p,k} \frac{ \langle \phi_p | d^{(1)}_k | \phi_v \rangle} { \epsilon_p - \epsilon_v} a_p^{\dagger} a_v
\end{eqnarray}
with $|\phi_i \rangle$ and $\epsilon_i$ are the single particle DHF wave function and energy of the the $i^{th}$ orbital, respectively. Here, $a,b$ denote for core orbitals, $p,q$ denote for virtual orbitals and $|\Phi_{ab\cdots}^{pq\cdots} \rangle = a_p^{\dagger} a_q^{\dagger} \cdots a_b a_a |\Phi_0 \rangle$.

Using the above DHF wave operators as the initial guess, the unperturbed perturbed wave operators in the MBPT method can be derived using the Bloch equation \cite{Lindgren1985}
\begin{eqnarray}
\left[\Omega^{(0)}_0,H_{DHF}\right]=\big(V_{res}\Omega^{(0)}_0 -\Omega^{(0)}_0 [V_{res} \Omega^{(0)}_0 ] )_{l}
\end{eqnarray}
and
\begin{eqnarray}
\left[\Omega_v^{(0)}, H_{DHF}\right]&=&\big( V_{res} (\Omega^{(0)}_0+\Omega^{(0)}_v) \nonumber\\
&& -\Omega^{(0)}_v [V_{res} (\Omega^{(0)}_0 +\Omega^{(0)}_v ] \big)_{l} , 
\end{eqnarray}
where `$l$' means that only the linked diagrams will contribute to the wave operator. In the MBPT($n=2,3$) approximations, $\Omega^{(0)}_{0/v}$ contains up to $n-1$ number of $V_{res}$. 

Similarly, the first-order perturbed wave operators in the MBPT method can be derived by extending the Bloch equation \cite{Singh2014, Chakraborty2023-2} as 
\begin{eqnarray}
 [\Omega_0^{(1)}, H_{DHF}] &=& (D \Omega_0^{(0)} +V_{res} \Omega_0^{(1)} )_l  
\end{eqnarray}
and
\begin{eqnarray}
 [\Omega_v^{(1)}, H_{DHF}] &=& \big(D (\Omega_0^{(0)} + \Omega_v^{(0)}) + V_{res} ( \Omega_0^{(1)} +\Omega_v^{(1)} )\big)_l \nonumber\\
&&  - \Omega_v^{(1)}  (V_{res} (\Omega^{(0)}_0 +\Omega^{(0)}_v) .  
\end{eqnarray}
It can be followed that $\Omega^{(1)}_{0/v}$ contains up to $n-2$ number of $V_{res}$ and one order of $D$ in the MBPT($n=2,3$) approximations.

\begin{figure}[t!]
\centering
\includegraphics[width=85mm,height=60mm]{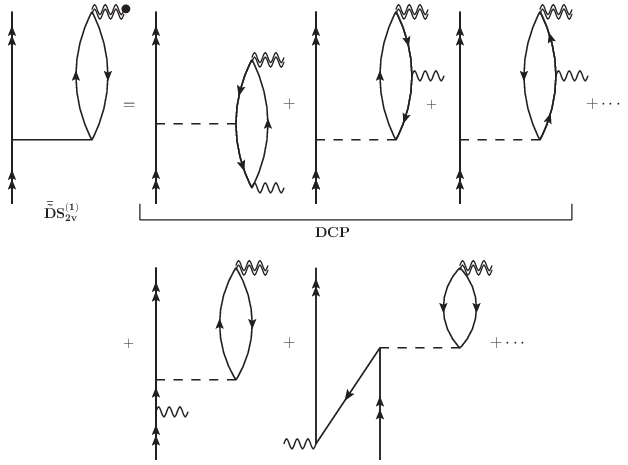}
\caption{Demonstration of breakdown of Goldstone diagram of the RCC term $\bar{\tilde{D}}S_{2v}^{(1)}$ as lower-order diagrams. All symbols used here stand for the same operators as in the previous figures.}
\label{s2v}
\end{figure}

In the RPA, $\Omega^{(0)}_{0/v}$ are equivalent to the DHF method and the first-order perturbed wave operators are obtained by 
\begin{eqnarray}
(H_{DHF} - {\cal E}_0) \Omega_0^{(1)} |\Phi_0 \rangle =  - D | \Phi_0 \rangle - U_0^{RPA}  |\Phi_0 \rangle  
\end{eqnarray}
and
\begin{eqnarray}
(H_{DHF} - {\cal E}_v) \Omega_v^{(1) } |\Phi_v \rangle =  - D | \Phi_v \rangle - U_v^{RPA}  |\Phi_v \rangle ,
\end{eqnarray}
where ${\cal E}_0=\langle \Phi_0 | H_{DHF} | \Phi_0 \rangle$, ${\cal E}_v=\langle \Phi_v | H_{DHF} | \Phi_v \rangle$ and the RPA potentials are defined as
\begin{eqnarray}
  U_0^{RPA} | \Phi_0 \rangle = \sum_{a,b} \left [ \langle b | V_{res} \Omega_0^{(1)} |b \rangle |a \rangle  -  \langle b | V_{res} \Omega_0^{(1)}  |a \rangle |b \rangle \right. \nonumber  \\ 
  \left. + \langle b | \Omega_0^{(1)\dagger} V_{res}  | b \rangle |a \rangle -  \langle b |\Omega_0^{(1)\dagger} V_{res}  |a \rangle |b \rangle  \right ]   \nonumber\\ 
\end{eqnarray}
and
\begin{eqnarray}
  U_v^{RPA} | \Phi_v \rangle = \sum_b \left [ \langle b | V_{res} \Omega_0^{(1)} |b \rangle |v \rangle   -  \langle b | V_{res} \Omega_v^{(1)}  |v \rangle |b \rangle \right. \nonumber  \\ 
  \left. + \langle b | \Omega_0^{(1)\dagger} V_{res}  | b \rangle |v \rangle  -  \langle b |\Omega_0^{(1)\dagger} V_{res}  |v \rangle |b \rangle  \right ] .  \nonumber\\
\end{eqnarray}

In the RCC theory {\it ansatz}, the unperturbed wave operators are defined as \cite{Lindgren1985, Cizek1969, Mukherjee1979}
\begin{eqnarray}
\Omega_0^{(0)} = e^{T^{(0)}}  
\end{eqnarray}
and
\begin{eqnarray}
\Omega_v^{(0)} = e^{T^{(0)}} S_v^{(0)}.
\label{eqrcc}
\end{eqnarray}
Extending these definitions to the first-order perturbed wave functions, we can define the corresponding wave operators as 
\begin{eqnarray}
\Omega_0^{(1)} = e^{T^{(0)}} T^{(1)} 
\end{eqnarray}
and
\begin{eqnarray}
\Omega_v^{(1)} = e^{T^{(0)}} \left ( S_v^{(1)} + S_v^{(0)}T^{(1)} \right ).
\end{eqnarray}
In the RCCSD method approximation, we define
\begin{eqnarray}
T^{(0)} &=&  T_{1}^{(0)} + T_{2}^{(0)}  \ \ \ \text{and} \ \ \ 
T^{(1)} =  T_{1}^{(1)} + T_{2}^{(1)}
\end{eqnarray}
and
\begin{eqnarray}
S_v^{(0)} &=&  S_{1v}^{(0)} + S_{2v}^{(0)}  \ \ \ \text{and} \ \ \
S_v^{(1)} =  S_{1v}^{(1)} + S_{2v}^{(1)},
\end{eqnarray}
where subscripts 1 and 2 denote single and double excitations, respectively. It yields
\begin{eqnarray}
\alpha_d^{S} &=& 
2\frac{\langle \Phi_v |\{ 1+ S_v^{(0)} \}^{\dagger} \bar{\tilde{D}}^S \{T^{(1)}(1+ S_v^{(0)}) + S_v^{(1)}\} |\Phi_v \rangle}{\langle \Phi_v | \{S_v^{(0)\dagger} +1 \} \bar{N} \{ 1+ S_v^{(0)} \} |\Phi_v \rangle} \nonumber \\  && + 2 \langle \Phi_0 | \bar{\tilde{D}}^S T^{(1)} |\Phi_0 \rangle 
\label{pols}
\end{eqnarray}
and
\begin{eqnarray}
\alpha_d^{T} &=&  2\frac{\langle \Phi_v |\{ 1+ S_v^{(0)} \}^{\dagger} \bar{\tilde{D}}^T \{T^{(1)}(1+ S_v^{(0)}) + S_v^{(1)}\} |\Phi_v \rangle}{\langle \Phi_v | \{S_v^{(0)\dagger} +1 \} \bar{N} \{ 1+ S_v^{(0)} \} |\Phi_v \rangle} , \nonumber \\ && 
\label{polt}
\end{eqnarray}
where $\bar{\tilde{D}}^{S/T}=e^{T^{(0)\dagger}}\tilde{D}^{S/T} e^{T^{(0)}}$ and $\bar{N}=e^{T^{(0)\dagger}}e^{T^{(0)}}$. 

In Eq. (\ref{pols}), the closed RCC term $\bar{\tilde{D}}^S T^{(1)}$ with respect to $|\Phi_0 \rangle$ contribute to $\alpha_c^{S}$ while the open part of RCC terms $\bar{\tilde{D}}^{S/T} T^{(1)}$ contribute to $\alpha_{cv}^{S/T}$.

\begin{table*}[t!]
\setlength{\tabcolsep}{3pt}
\caption{Calculated static scalar and tensor polarizability values of the ground and excited states of Cs atom using the DHF and many-body methods. All values are in a.u.. }
\centering
\begin{tabular}{ll rrrrr c ccccc}
\hline\hline\\
&&\multicolumn{5}{c}{$\alpha_d^{S}$}&&\multicolumn{5}{c}{$\alpha_d^{T}$}\\
\cline{3-7}\cline{9-13}
State&&DHF&MBPT(2)&RPA&MBPT(3)&RCCSD&&DHF&MBPT(2)&RPA&MBPT(3)&RCCSD\\
\hline
6$S_{1/2}$&&664.6&571.3&625.2&544.2&405.0&&  &  &  & & \\[1ex]
6$P_{1/2}$&&1379.7&1305.5&1348.2&1732.5&1330.9&&  &  &  & & \\[1ex]
6$P_{3/2}$&&1620.1&1536.3&1584.3&2053.4&1638.5&&$-215.8$&$-234.5$&$-224.5$&$-422.9$&$-261.4$\\[1ex]
5$D_{3/2}$&&151.5&187.3&169.3&78.4&$-342.5$ &&257.8&212.6&237.7&85.4&360.7\\[1ex]
5$D_{5/2}$&&43.7&82.6&62.7&64.1&$-443.1$&&513.1&444.7&482.8&163.5&677.1\\[1ex]
7$S_{1/2}$&&8084.5&7891.4&8008.2&7983.7&6197.7&&  &  &  & &\\[1ex]
7$P_{1/2}$&&21644.8&21394.0&21543.7&36205.3&29470.1&&  &  &  & & \\[1ex]
7$P_{3/2}$&&25808.2&25527.5&25694.0&43307.6&36888.3&&$-3038.8$&$-3067.1$&$-3052.9$&$-6359.7$&$-4385.1$\\[1ex]
8$S_{1/2}$&&47192.9&46849.8&47063.0&49332.3&38020.8&&  &  &  & &\\[1ex]
8$P_{1/2}$&&147930.1&147402.8&147732.8&266682.6&219973.3&&  &  &  & &\\[1ex]
8$P_{3/2}$&&177595.6&177004.4&177372.0&322044.1&278993.8&&$-19460.6$&$-19497.1$&$-19481.6$&$-42552.2$&$-30381.1$\\[1ex]
9$S_{1/2}$&& 187510.5&186960.9&187311.5&202567.0&153252.1&&  &  &  & & \\[1ex]
9$P_{1/2}$&&647919.8&646990.2&647600.9&1217278.5&1004191.9&&  &  &  & & \\[1ex]
9$P_{3/2}$&&781171.4&780127.8&780809.3&1478366.3&1282853.7&&$-80007.7$&$-80048.7$&$-80037.0$&$-184328.1$&$-132655.0$\\[1ex]
\hline\hline\\
\end{tabular}
\label{tab1}
\end{table*}

In order to corroborate our earlier statements that the RCC method, even at the RCCSD approximation, includes CP, PC, DCP and their correlations to all-orders, we demonstrate this by expressing the RCC terms in Goldstone diagrams and their break-downs into MBPT diagrams. In Fig. \ref{cor1}, we show core contributing Goldstone diagram for $(\bar{\tilde{D}}T_1^{(1)})_c$ and its breakdown in terms of MBPT diagrams. We have also shown the open diagrams from the term $(\bar{\tilde{D}}T_1^{(1)})_{cv}$ contributing to core-valence sector of the the $\alpha_d^{S/T}$ values in Fig. \ref{corv}. We have also classified these diagrams under DHF, RPA types corresponding to CP contributions and non-RPA types corresponding to PC contributions. This demonstrates that the core and core-valence contributing RCC terms take into account CP, PC and their correlations to all-orders. It can be shown that contributions from $\bar{\tilde{D}}T_2^{(1)}$ corresponds to DCP and other non-CP and non-PC effects. Similarly, it is evident from Fig. \ref{s1v} that the Goldstone diagram representing the RCC term $\bar{\tilde{D}}S_{1v}^{(1)}$ that takes into account contributions from the single excitations to $\alpha_{v}^{S/T}$ also contains the CP, PC and their correlations involving the valence electron to all-orders. The Goldstone diagram shown in Fig. \ref{s2v} for $\bar{\tilde{D}}S_{2v}^{(1)}$ represents for the double excitation contributions to $\alpha_{v}^{S/T}$. Its breakdown in terms of MBPT diagrams show that it includes contributions from DCP and many non-CP and non-PC effects that cannot be taken into account in the sum-over-states approach. Since amplitude solving equations for the single and double excitation operators in the RCC method are coupled, all these terms are correlated through the LR approach in the determination of the $\alpha_d^{S/T}$ values. Moreover, core and valence correlation contributions are also coupled in the LR approach of the RCC method.

\begin{figure*}[t!]
\setlength{\tabcolsep}{10pt}
\centering
\begin{tabular}{c c}\\
\includegraphics[width=80mm,height=48mm]{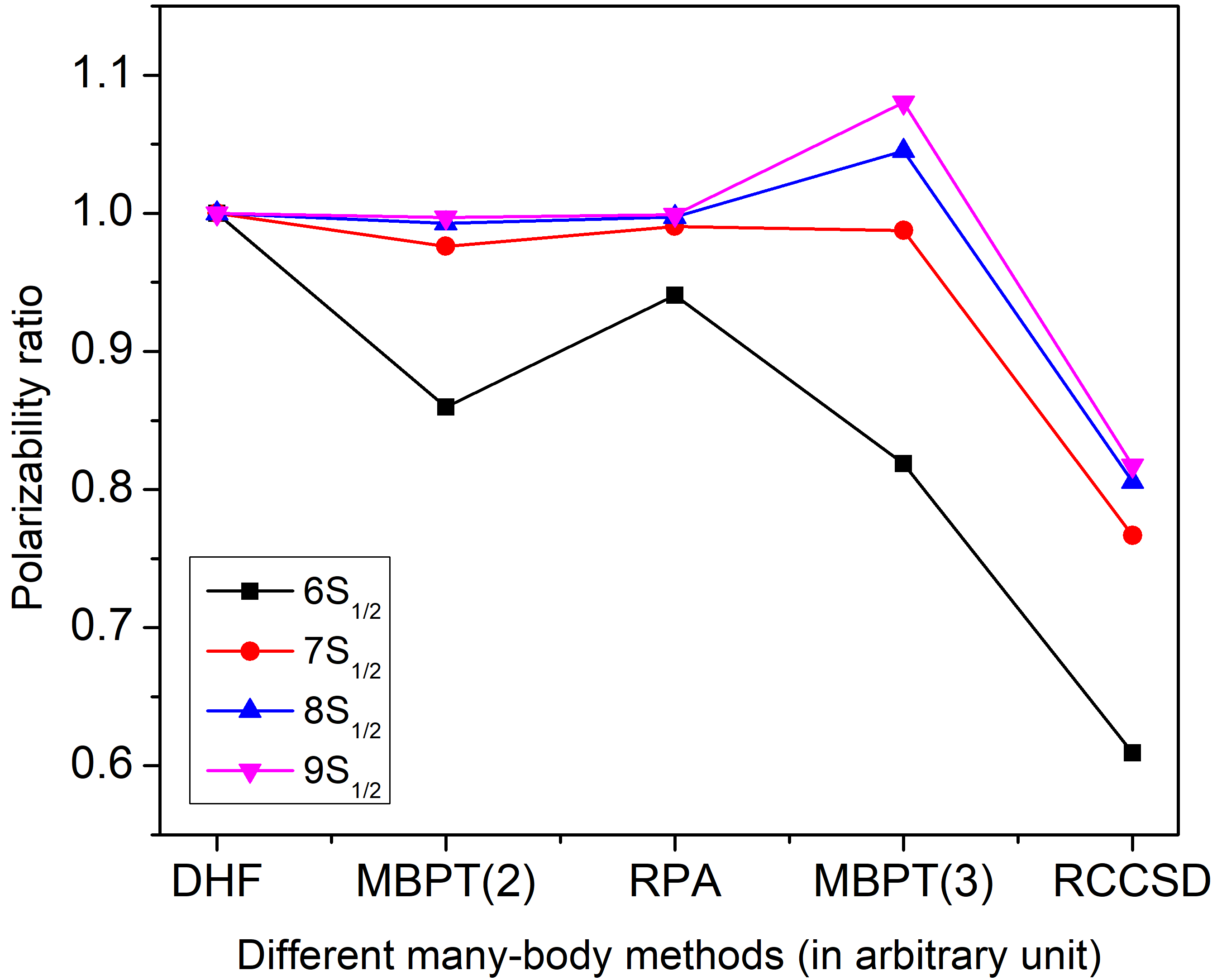} &
\includegraphics[width=80mm,height=48mm]{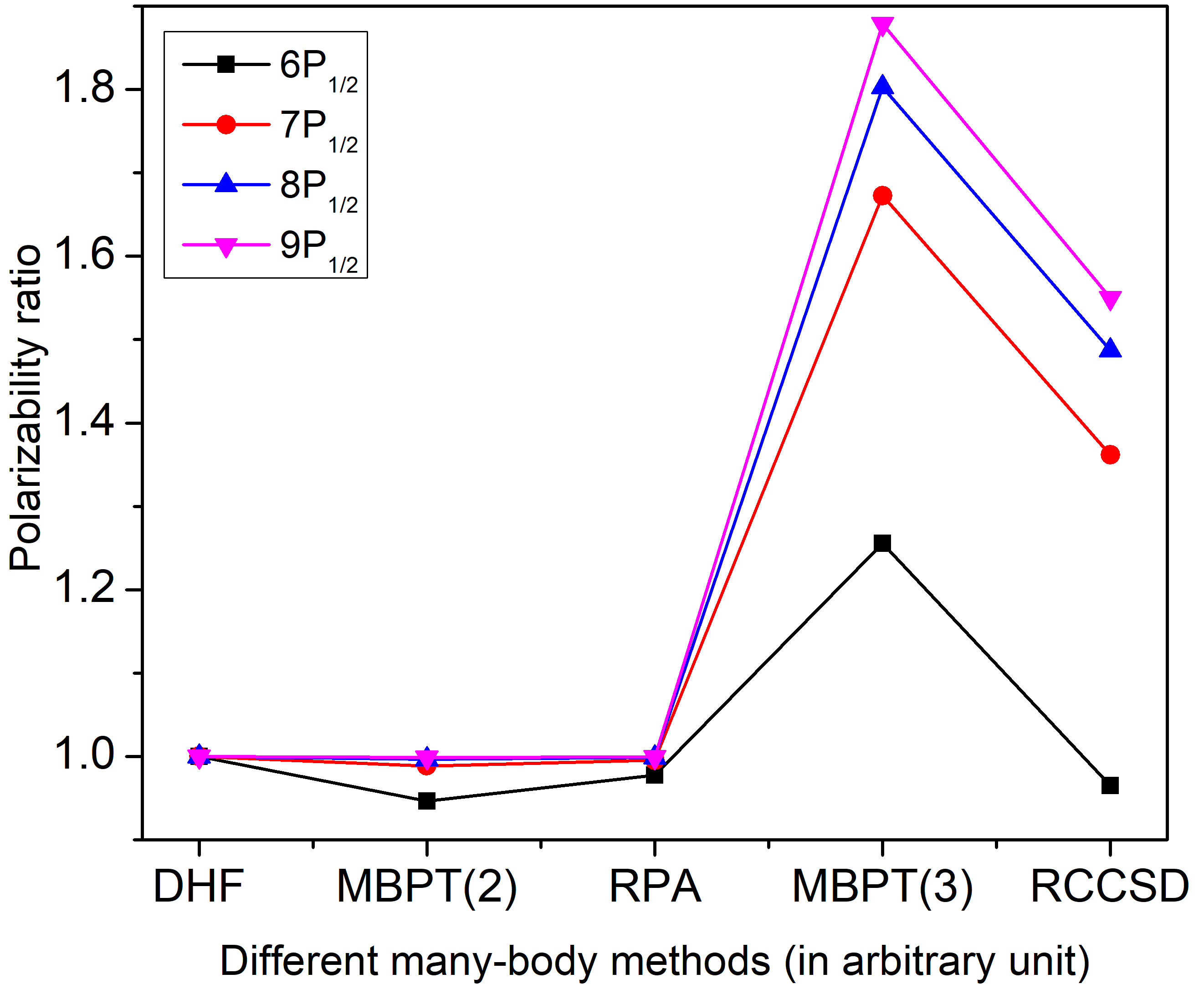}\\ 
(a) Correlation trends of $\alpha_d^S$ in the $nS_{1/2}$ states & (b) Correlation trends of $\alpha_d^S$ in the $nP_{1/2}$ states  \\[5ex]
\includegraphics[width=80mm,height=48mm]{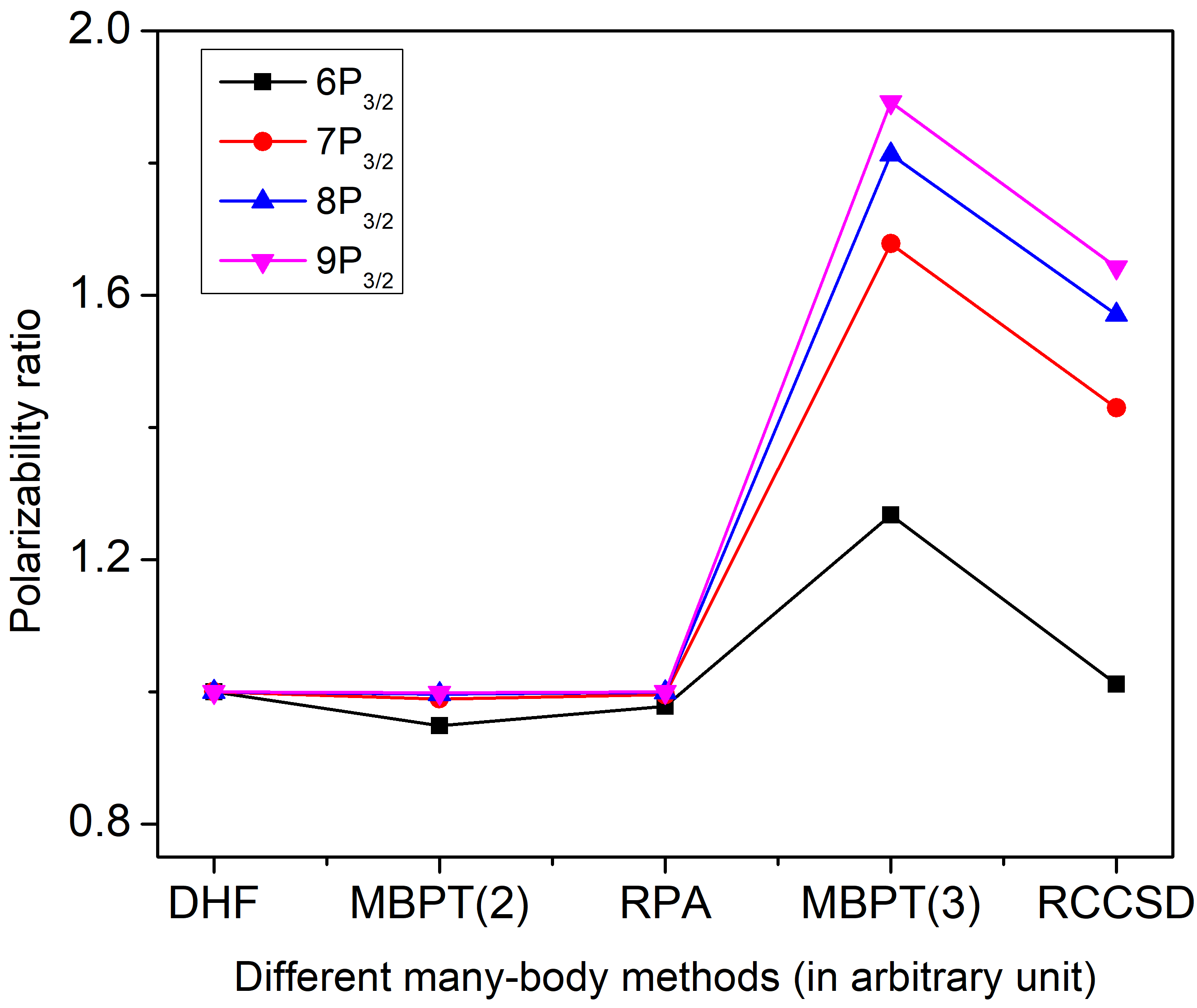} &
\includegraphics[width=80mm,height=48mm]{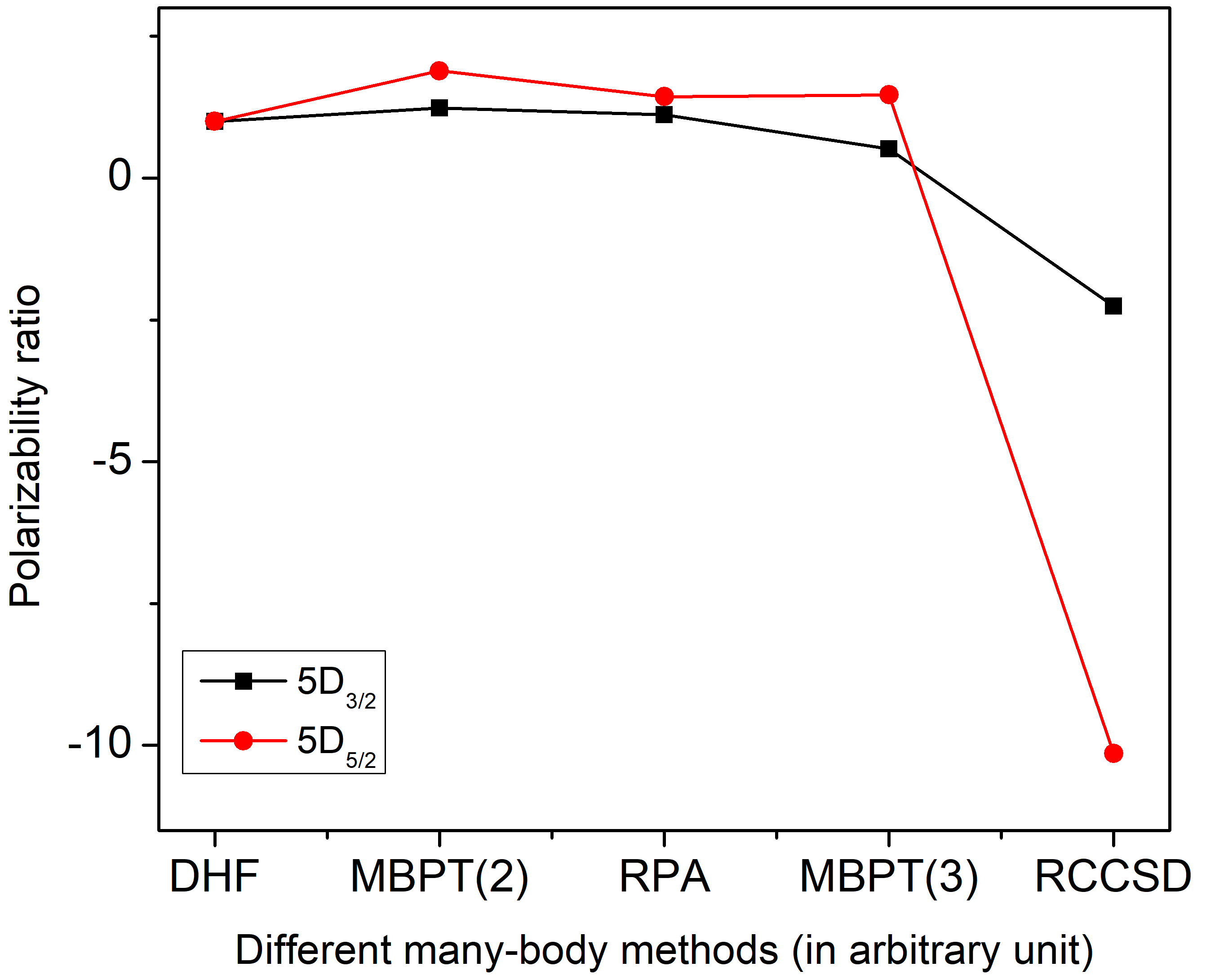}\\ 
(c) Correlation trends of $\alpha_d^S$ in the $nP_{3/2}$ states & (d) Correlation trends of $\alpha_d^S$ in the $5D_{3/2,5/2}$ states \\[5ex]
\end{tabular}
\caption{Ratios of scalar polarizability values from different many-body methods and their DHF values. These plots demonstrate the amount of electron correlation effects captured in the many-body methods in the determination of $\alpha_d^S$ of different states in the Cs atom.}
\label{fig1}
\end{figure*} 

\section{Results \& Discussion}\label{secre}

For accurate calculations of wave functions of all the considered states of Cs, we have employed an extensive size basis set of primitive Gaussian-type orbitals (GTOs), specifically tailored to capture the electronic correlations more effectively. The basis set consisted of 40, 39, 38, 37, 36, 35, and 34 GTOs for the $s, p, d, f, g, h$, and $i$ symmetries, respectively. This large basis set is expected to ensure that the calculations are both comprehensive and accurate, accounting for a wide range of radial part in the atomic wave function. As mentioned earlier, we have employed the DHF, MBPT(2), MBPT(3), RPA and RCCSD methods to calculate values for $\alpha_d^S$ and $\alpha_d^T$ of the Cs atom. The results are presented in Table \ref{tab1}. As shown in this table, the differences between the DHF and MBPT(2) values of $\alpha_d^{S/T}$ are small, indicating contributions from the lowest-order CP effects are minimal. Now, comparing the results from the MBPT(2) and RPA methods, we again find a very small difference in the polarizability values. However, the RPA results are much more closer to the DHF values than those from the MBPT(2) method. As the RPA approach includes CP effects to all-orders, this small difference indicates that there are cancellations among the lowest-order and higher-order CP effects in the evaluation of the $\alpha_d^{S/T}$ values.  

Now we compare the $\alpha_d^S$ results from the MBPT(2) and RPA methods with MBPT(3) values. Other than $S_{1/2}$ states, we see very large differences in the polarizability values; especially for the $D_{3/2}$ and $D_{5/2}$ states. This shows that the polarizability values are very much sensitive to the PC correlation effects than the CP effects. The dominance of PC over CP effects are much more prominent in the RCCSD method. In the $5D$ states, the all-order PC effects flip the sign of the scalar polarizability values. A similarly large difference can be observed between MBPT(2) and MBPT(3) values for $\alpha_d^T$. However, it is interesting to note that the RCCSD values are closer to the MBPT(2) or RPA results than the MBPT(3) method, suggesting that there are huge cancellations among the lowest-order PC effects and all-order PC effects.

\begin{figure*}[t!]
\setlength{\tabcolsep}{10pt}
\centering
\begin{tabular}{c c}\\
\includegraphics[width=80mm,height=48mm]{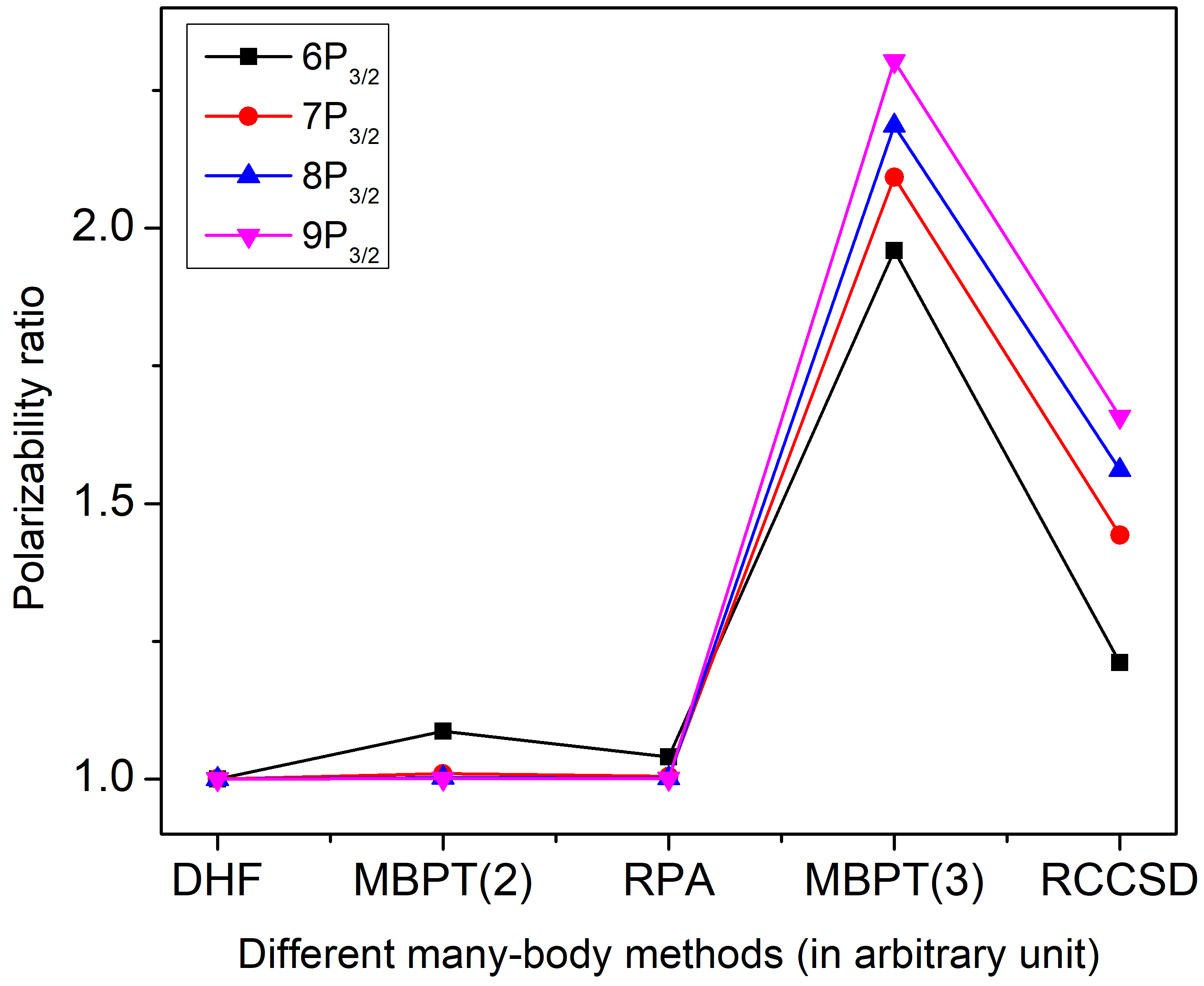} &
\includegraphics[width=80mm,height=48mm]{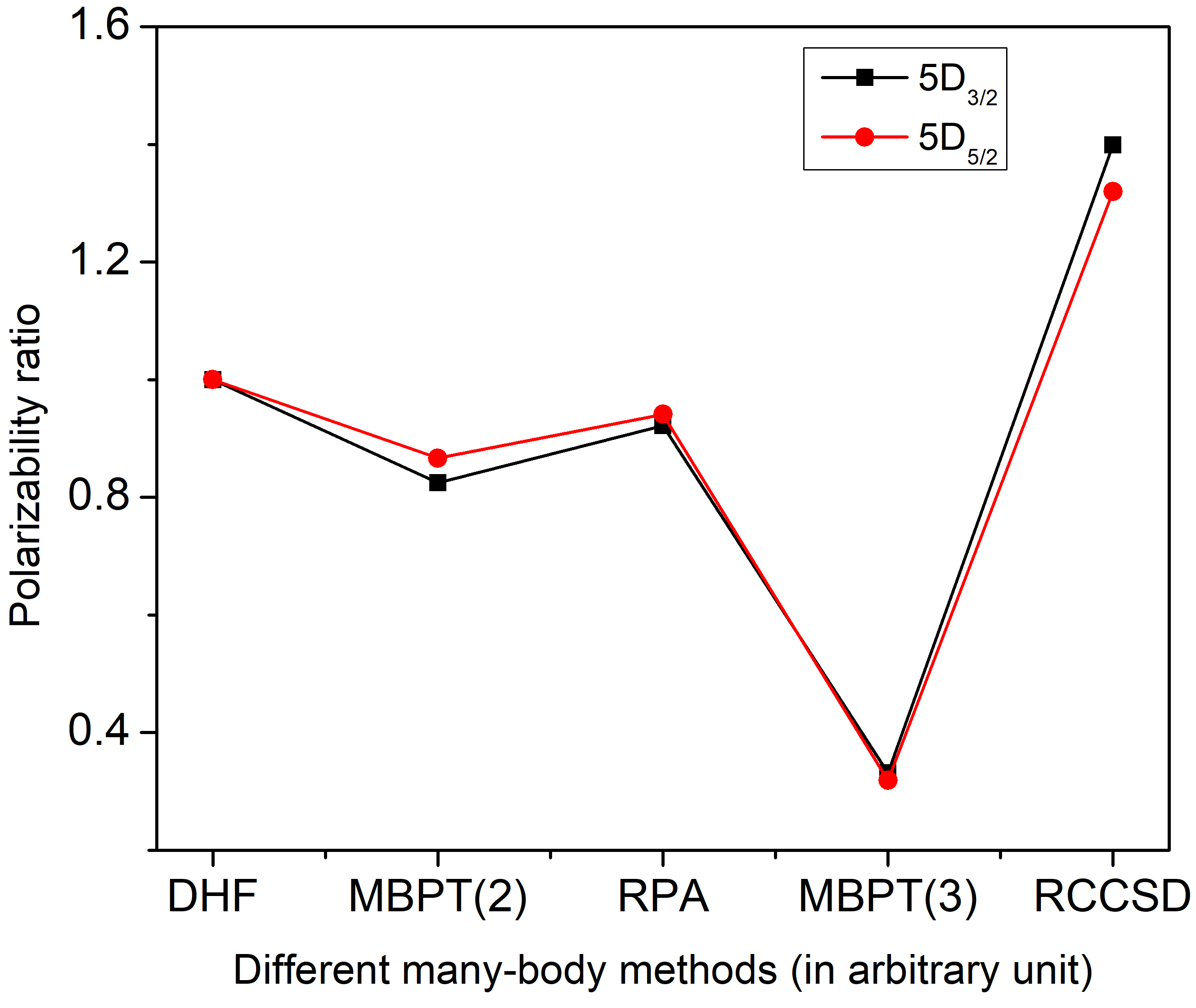}\\ 
(a) Correlation trends of $\alpha_d^T$ in the $nP_{3/2}$ states & (b) Correlation trends of $\alpha_d^T$ in the $5D_{3/2,5/2}$ states  \\[5ex]
\end{tabular}
\caption{Ratios of tensor polarizability values from different many-body methods and their DHF values. The figures highlight how effectively electron correlation effects are captured by various many-body approaches in the computation of the $\alpha_d^T$ values for different states of the Cs atom.}
\label{fig2}
\end{figure*}

To gain a more comprehensive understanding of the contributions from different many-body methods in estimating the $\alpha_d^S$ and $\alpha_d^T$ of different states in Cs, we present ratios of these values from each method to the corresponding DHF values. Fig. \ref{fig1} illustrates the trends in the scalar polarizability values, while Fig. \ref{fig2} provides the corresponding trends for tensor polarizabilities. The primary factors influencing these calculations are the non-RPA effects, which dominate the correlation trends. To explore this further, we analyzed the correlation trends among the atomic states belonging to the same angular momentum symmetry for the scalar polarizability values $\alpha_d^S$. As shown in Fig. \ref{fig1}, the correlation trends from different $S_{1/2}$ states are not the same. This implies that states with the same angular momentum are not necessarily influenced by the correlation effects in the similar manner. For the ground state, the cancellation between lowest-order and all-order CP effects is particularly strong. The PC effects further decrease the polarizability value in the MBPT(3) and RCCSD methods. However, for other $S_{1/2}$ states, $\alpha_d^S$ is relatively insensitive to the CP effects. Notably, the lowest-order non-RPA effects from the MBPT(3) method do not contribute to $\alpha_d^S$ of the $7S_{1/2}$ state, however, for the $8S_{1/2}$ and $9S_{1/2}$ states, these effects increase the polarizability values. Once again, the all-order RCCSD calculations decrease the values. 

For the $nP_{1/2}$ and $nP_{3/2}$ states, similar trends are observed in the correlation plots. Unlike the $S_{1/2}$ states, where non-RPA effects decrease the values of $\alpha_d^S$, the non-RPA effects in the $P_{1/2}$ and $P_{3/2}$ states actually increase the overall values of $\alpha_d^S$. However, as can be seen in the plots, for the $6P_{1/2}$ and $6P_{3/2}$ states, the polarizability values are more sensitive to the CP effects than the other $P_{1/2,3/2}$ states. Another interesting thing to note is that the PC effects are less dominating in the $6P$ states compared to other $nP$ states. In all the $P_{1/2,3/2}$ states, the lowest-order non-RPA effects increase the polarizabilities at the MBPT(3) level. Subsequently, the all-order RCCSD contributions reduce the polarizability values again. This trend highlights the complex roles of non-RPA effects and their importance in accurately capturing the behavior of the correlation effects in the excited states. The effects of the non-RPA corrections are significantly pronounced in the $5D_{3/2,5/2}$ states. In these states, changes in the polarizability values from the DHF method to the MBPT(3) method is small. However, all-order PC effects from the RCCSD method are very dominating, decreasing the polarizability values and even flipping their signs. The differences between the MBPT(3) and RCCSD values for the scalar polarizabilities of the $5D$ states exceeds 100\%, emphasizing the critical roles played by the all-order effects captured in the RCCSD method. These substantial deviations underscore the importance of including higher-order corrections, as they significantly alter the polarizability values compared to the lower-order perturbative methods, highlighting the necessity of better all-order treatments of correlation effects in the accurate predictions of polarizability values; particularly in Cs.

Next, we focus on the correlation trends of the tensor polarizability values for the $P_{3/2}$ and $D_{3/2,5/2}$ states. As shown in Fig. \ref{fig2}, all the $nP_{3/2}$ states exhibit a similar correlation trend, where the non-RPA effects increase the magnitudes of the polarizability values. Like in the case of $\alpha_d^S$, here also correlations in the $6P_{3/2}$ state is more sensitive to the CP effects than the other $nP_{3/2}$ states. A particularly interesting trend is observed for the $5D$ states. In these states, the MBPT(3) contributions reduce the tensor polarizabilities compared to the RPA level. However, the all-order contributions from the RCCSD method reverse the effect and increase the results again. The differences between the MBPT(3) and RCCSD values exceed 70\%, highlighting the importance of the higher-order effects in the $\alpha_d^{T}$ values that can only be captured by the RCCSD method.

\begin{table*}[t]
\setlength{\tabcolsep}{3pt}
\centering
\caption{Contributions from different RCC terms to the polarizability values of $\alpha_d^S$ and $\alpha_d^T$ for the considered states of the Cs atom. Terms with subscripts $`c$' and $`cv$' correspond to core and core-valence contributions, respectively. Contributions given under `Norm' represent corrections to the results due to normalization factors of the wave functions. Contributions from other non-linear terms of the RCCSD method are given together under ``Others".}
\begin{tabular}{cc cccc rr rr}
\hline\hline
Polarizability&State&\multicolumn{8}{c}{Contributions}\\
\cline{3-10}\\
&&$(\bar{\tilde{D}}T_1^{(1)}+c.c.)_{c}$&$(\bar{\tilde{D}}T_1^{(1)}+c.c.)_{cv}$&$\bar{\tilde{D}}S_{1v}^{(1)}+c.c.$&$\bar{\tilde{D}}S_{2v}^{(1)}+c.c.$&Norm& Others& Breit & QED\\
\hline
$\alpha_d^S$&6$S_{1/2}$&15.42 &$-0.62$ &456.29 &$-11.46$ &$-9.22$ &$-45.76$ &$-0.10$&0.45\\[1ex]
&6$P_{1/2}$&15.42 &$\sim$0.0&1442.78 &$-20.34$ &$-18.79$ &$-95.57$ &7.83&$-0.41$\\[1ex] 
&6$P_{3/2}$&15.42 &$\sim$0.0 &1761.39 &$-24.23$ &$-20.12$ &$-101.70$ &7.62&0.08\\[1ex]
&5$D_{3/2}$&15.42 &$-0.30$ &$-356.99$ &9.87 &27.44 &$-31.74$ &$-6.01$&$-0.20$ \\[1ex]
&5$D_{5/2}$&15.42 &$-0.38$ &$-485.69$ &10.71 &31.92 &$-7.94$ &$-6.81$&$-0.35$\\[1ex]
&7$S_{1/2}$&15.42 &$-0.11$ &7035.08 &$-29.44$ &$-145.45$ &$-682.49$ &$-2.56$&7.25\\[1ex]
&7$P_{1/2}$&15.42 &$\sim$0.0 &30358.97 &$-66.18$ &$-476.06$ &$-615.94$ &254.77&$-0.90$\\[1ex]
&7$P_{3/2}$&15.42 &$\sim$0.0 &37804.60 &$-79.86$ &$-514.28$ &$-658.92$ &308.60&12.71\\[1ex]
&8$S_{1/2}$&15.42 &$-0.04$ &43885.92 &$-55.17$ &$-850.31$ &$-5002.73$ &$-24.51$&52.18\\[1ex]
&8$P_{1/2}$&15.42 &$\sim$0.0 &225282.56 &$-124.65$ &$-3265.82$ &$-4048.54$ &2103.06&11.28\\[1ex]
&8$P_{3/2}$&15.42 &$\sim$0.0 &283288.74 &$-148.97$ &$-3558.05$ &$-3414.08$ &2688.23&122.47\\[1ex]   
&9$S_{1/2}$&15.42 &$-0.02$ &180432.14 &$-88.40$ &$-3358.39$ &$-23880.17$ &$-127.63$&259.12\\[1ex]
&9$P_{1/2}$&15.42 &$\sim$0.0 &1026237.23 &$-200.67$ &$-14822.14$ &$-16682.67$ &9516.60&128.11\\[1ex]
&9$P_{3/2}$&15.42 &$\sim$0.0 &1299488.89 &$-241.43$ &$-16107.75$ &$-13211.79$ &12309.27&601.09 \\[1ex]
\hline\\
$\alpha_d^T$&6$P_{3/2}$&0.0 &$\sim$0.0 &$-303.50$ &$-0.97$ &3.16 &40.57 &$-0.92$&0.28\\[1ex]
&5$D_{3/2}$&0.0 &0.18 &411.54 &$-4.20$ &$-27.84$ &$-22.33$ &3.33&$-0.02$\\[1ex]
&5$D_{5/2}$&0.0 &0.38 &794.58 &$-6.77$ &$-47.45$ &$-68.96$ &5.01&0.26\\[1ex]
&7$P_{3/2}$&0.0&$\sim$0.0 &$-4929.15$ &$-0.21$ &61.27 &513.68 &$-34.10$&3.37\\[1ex]
&8$P_{3/2}$&0.0 &$\sim$0.0 &$-33599.38$ &0.52 &393.31 &3097.86 &$-292.81$&19.37\\[1ex]            
&9$P_{3/2}$&0.0 &$\sim$0.0 &$-145747.83$ &2.20 &1685.04 &12659.50 &$-1315.23$&61.34\\[1ex]            
\hline\hline
\end{tabular}
\label{tab3}
\end{table*}

To explore contributions of different correlation effects to the values of $\alpha_d^S$ and $\alpha_d^T$ through different RCC terms, we present results from these terms in Table \ref{tab3}. As it was mentioned earlier, the closed part of $\bar{\tilde{D}}T_1^{(1)}$ and its complex conjugate (c.c.) term, $T_1^{(1)\dagger}\bar{\tilde{D}}$, correspond to the core correlations, while their open parts contribute to the core-valence correlations. In the table, these contributions are listed separately under the terms $(\bar{\tilde{D}}T_1^{(1)} + c.c.)_{c}$ for core correlations and $(\bar{\tilde{D}}T_1^{(1)} + c.c.)_{cv}$ for core-valence correlations. The terms involving $S_v^{(0)}/S_v^{(1)}$ represent the valence correlations. It is evident from this table that the dominant contributions to both $\alpha_d^S$ and $\alpha_d^T$ arise from the $\bar{\tilde{D}}S_{1v}^{(1)}$ and its c.c. terms. The core correlations originating from $\bar{\tilde{D}} T_1^{(1)}$ and c.c. terms encompass contributions from both the singly and doubly excited configurations. These terms account not only for the core contributions in the RPA method but also for the PC contributions to the core correlations in the MBPT(3) method to all-orders as shown in the figures of the previous section. 

\begin{table*}[t]
\setlength{\tabcolsep}{5pt}
\centering
\caption{ List of recommended values for $\alpha_d^S$ and $\alpha_d^T$ (in a.u.) from our calculations. We also compare our results with the other theoretical and available experimental values.}
\begin{tabular}{l rrr crrr}
\hline\hline\\
State & \multicolumn{3}{c}{$\alpha_d^S$} && \multicolumn{3}{c}{$\alpha_d^T$} \\
\cline{2-4}\cline{6-8}\\
& This work & Theory & Experiment && This work & Theory & Experiment\\
\hline
6$S_{1/2}$&405(5)&399.8 \cite{Safronova1999}&401.00(6) \cite{Amini2003}&&  & & \\
& &401.0(6) \cite{Chakraborty2023}& &&  & & \\
& &406(8) \cite{Safronova2016}& &&  & & \\
& &398.4(7) \cite{Iskrenova2007}& &&  & & \\
& &396.32 \cite{Tang2014}& &&  & & \\[2ex]

6$P_{1/2}$&1331(18)&1339(43) \cite{Safronova2016}&1371(9) \cite{Hunter1988}&&  & & \\
& &1338(54) \cite{Iskrenova2007}&1328.35(12) \cite{Hunter1992}&&  & & \\
& &1332.2 \cite{Tang2014}&&&  & & \\[2ex]

6$P_{3/2}$& 1638(20) &1651(46) \cite{Safronova2016}&1665(13) \cite{Hunter1988}&& $-261.4(4.8)$ &$-260(11)$ \cite{Safronova2016}&$-262.4(1.5)$ \cite{Tanner1988} \\
& &1648(58) \cite{Iskrenova2007}&1640(2) \cite{Tanner1988}&&  &$-261(13)$ \cite{Iskrenova2007}& \\
& &1644.7 \cite{Tang2014}&&&  &$-262.95$ \cite{Tang2014}& \\[2ex]

5$D_{3/2}$&$-343(20)$&$-335(38)$ \cite{Safronova2016}& &&360.7(6.9)& 357(25) \cite{Safronova2016}& \\
& &$-352(69)$ \cite{Iskrenova2007}& && &370(28) \cite{Iskrenova2007}& \\
& &$-346.30$ \cite{Tang2014}&&&  &359.26 \cite{Tang2014}& \\[2ex]

5$D_{5/2}$&$-443(23)$&$-439(42)$ \cite{Safronova2016}& &&677(13)&677(34) \cite{Safronova2016}& \\
& &$-453(70)$ \cite{Iskrenova2007}& &&  &691(40) \cite{Iskrenova2007}& \\
& &$-450.30$ \cite{Tang2014}&&&  &680.47 \cite{Tang2014}& \\[2ex]

7$S_{1/2}$&6198(41)&6237(42) \cite{Safronova2016}&6238(6) \cite{Bennett1999}&&  & & \\
& &6238(41) \cite{Iskrenova2007}&6207.9(2.4) \cite{Quirk2024}&&  & & \\
& &6242.5 \cite{Tang2014}&&&  & & \\
& &6208 \cite{Bouchiat1983}&&&  & & \\[2ex]

7$P_{1/2}$&29470(350)&29880(160) \cite{Safronova2016}&29660(50) \cite{Toh2014}&&  & & \\
& &29890(700) \cite{Iskrenova2007}&29500(600) \cite{Domelunksen1983}&&  & & \\
& &30138 \cite{Tang2014}&&&  & & \\[2ex]

7$P_{3/2}$&36888(432) &37510(170) \cite{Safronova2016}&37280(70) \cite{Toh2014}&&  $-4385(69)$&$-4408(50)$ \cite{Safronova2016}&$-4413(29)$ \cite{Toh2014}\\
& &37500(800) \cite{Iskrenova2007}&37800(800) \cite{Domelunksen1983}&&  &$-4410(170)$ \cite{Iskrenova2007}&$-4420(120)$ \cite{Domelunksen1983}\\
& &37867 \cite{Tang2014}&&&  &$-4431.9$ \cite{Tang2014}& \\[2ex]

8$S_{1/2}$&38021(235)&38270(260) \cite{Safronova2016}&38370(380) \cite{Weaver2012} &&  & & \\
&&38270(280) \cite{Iskrenova2007}&38260(290) \cite{Gunawardena2007}&&  & & \\
& & & 38110(50) \cite{Antypas2011}&&  & & \\[2ex]

8$P_{1/2}$&219973(2488)&223300(1400) \cite{Safronova2016}& &&  & & \\
& &223000(2000) \cite{Iskrenova2007}& &&  & & \\[2ex]

8$P_{3/2}$& 278994(3180) &284500(1700) \cite{Safronova2016}& && $-30381(438)$ &$-30570(410)$ \cite{Safronova2016}& \\
& &284000(3000) \cite{Iskrenova2007}& &&  &$-30600(600)$ \cite{Iskrenova2007}& \\[2ex]

9$S_{1/2}$&153252(1152)&153700(1000) \cite{Safronova2016}&150700(1500) \cite{Weaver2012} &&  & & \\
& &153700(1700) \cite{Iskrenova2007}& &&  & & \\[2ex]

9$P_{1/2}$& 1004192(9781) &1021400(5600) \cite{Safronova2016}& &&  & & \\
& &1021000(7000) \cite{Iskrenova2007}& &&  & & \\[2ex]

9$P_{3/2}$& 1282854(12283) &1312900(7000) \cite{Safronova2016}& && $-132655(1367)$ &$-134700(1700)$ \cite{Safronova2016}& \\
& &1312000(7000) \cite{Iskrenova2007}& &&  &$-135000(2000)$ \cite{Iskrenova2007}& \\[2ex]
\hline\hline      
\end{tabular}
\label{tab4}
\end{table*}

Similarly, the valence correlation contributions from the RPA are captured by the $\bar{\tilde{D}} S_{1v}^{(1)} + S_{1v}^{(1)\dagger} \bar{\tilde{D}}$ terms in the RCCSD method, which include contributions from the PC correlations and higher-order effects involving the valence electron \cite{Sahoo2007, Sahoo2009}. Another significant RCCSD term is $\bar{\tilde{D}} S_{2v}^{(1)} + c.c.$, whose contributions cannot be neglected for accurate determination of the polarizabilities. For example, in the case of the ground state of the Cs atom, these terms contribute up to 3\% of the total value. Additionally, we present corrections to the polarizability values due to wave function normalization, labeled as ‘Norm’. As the table indicates, contributions from ‘Norm’ cover 1-2\% of the polarizability values, highlighting their significance. The correlation contributions to $\alpha_d^S$ and $\alpha_d^T$ arising from the other RCCSD terms, such as $\bar{\tilde{D}}T_{1/2}^{(1)}S_{1/2v}^{(0)}$, $T_{1/2}^{(1)\dagger} \bar{\tilde{D}} S_{1/2v}^{(0)}$, $S_{1/2v}^{(0)\dagger} \bar{\tilde{D}} S_{1/2v}^{(1)}$ and $S_{1/2v}^{(1)\dagger} \bar{\tilde{D}} S_{1/2v}^{(0)}$ are also non-negligible. These are basically contributions from the non-RPA effects, many of which cannot be considered as a part of the PC correlations. We present the contributions from these nonlinear terms under the label ‘Others’ in the table above. Inclusion of these contributions is also important and can contribute up to 10-15\% of the total values for scalar polarizabilities of the $S_{1/2}$ and $D_{3/2}$ states. For the $P_{1/2,3/2}$ states, these terms also contribute about 1-7\% of their total values. Similar trends are observed in the tensor polarizability calculations as well. In the above table, we present contributions from corrections arising from the Breit interaction and QED effects. As can be seen in the table, the Breit contributions substantially influence the polarizability values. These contributions are particularly prominent in the $5D_{3/2,5/2}$ states, where they account for about 2\% of the total values. For other states, the Breit contributions are about 1\% and cannot be neglected. In contrast, the QED contributions are only about 0.1\% of the total values, making their impact less on the polarizability estimations. It is worth noting at this stage that we only make ballpark estimation of QED contributions by using model potentials for the lowest-order vacuum-polarization and self-energy correction terms through the atomic Hamiltonian. However, these estimations can have 100\% uncertainties. 

The recommended values for $\alpha_d^S$ and $\alpha_d^T$ are given as the values obtained from the RCCSD method along with other corrections. These values are presented in Table \ref{tab4}, along with the estimated uncertainties, which are derived from the leading order triple excitations and estimated QED effects. In the table, we compare our results with other recent high-precision relativistic calculations and experimental data \cite{Safronova2004, Safronova1999, Chakraborty2023, Iskrenova2007, Safronova2016, Tang2014, Bouchiat1983, Amini2003, Bennett1999, Quirk2024, Weaver2012, Gunawardena2007, Antypas2011, Hunter1988, Hunter1992, Toh2014, Domelunksen1983, Tanner1988}. As observed in the table, our recommended value for the $6S_{1/2}$ state is in good agreement with the experimental value when the error bar is taken into account. It is worth noting that this value can be further refined and the uncertainty minimized by including full triple excitations in the RCC calculation, which would improve the precision of the results. A preliminary investigation shows triple excitations help bring this value very close to the experimental result. For the $6P_{1/2,3/2}$ states, our recommended values for $\alpha_d^S$ are more accurate compared to other theoretical estimates. Moreover, our results for the $6P_{1/2}$ and $6P_{3/2}$ states align more closely with the experimental values from Refs. \cite{Hunter1992} and \cite{Tanner1988}, respectively. This improvement represents a significant step forward in achieving more accurate and reliable values for $\alpha_d$ in the heavier atomic systems. Our recommended value for $\alpha_d^T$ of the $6P_{3/2}$ state is also in good agreement with the experimental value reported in Ref. \cite{Tanner1988}. This agreement reinforces the reliability of our method for accurate calculations of $\alpha_d$ in the LR approach. 

For the $5D_{3/2}$ and $5D_{5/2}$ states, no experimental values are currently available, so we have turned to comparisons with theoretical results from other recent works \cite{Safronova2016, Iskrenova2007, Tang2014}. Notably, our results for both the scalar and tensor polarizabilities of these states are in excellent agreement with the values obtained using the sum-over-states approach. Our recommended result for the $7S_{1/2}$ state is closer to the experimental value reported in Ref. \cite{Quirk2024} than the value from Ref. \cite{Bennett1999}. For the $7P_{1/2}$ and $7P_{3/2}$ states, our estimated values are also well within the experimental uncertainties. For the $8S_{1/2}$ state, our reported value for the scalar polarizability is in better agreement with the most precise experimental value available from Ref. \cite{Antypas2011}. In the case of the $8P_{1/2}$ and $8P_{3/2}$ states, no experimental values are currently available. Therefore, we once again compare our results with the sum-over-states results from other theoretical studies. As shown in the table, our results are about 1-2\% smaller than those reported in Refs. \cite{Safronova2016, Iskrenova2007}. For the $9S_{1/2}$ state, our result is much closer to the experimental value reported in Ref. \cite{Weaver2012} than to the other theoretical estimations. In contrast, for the scalar and tensor polarizabilities of the $9P_{1/2}$ and $9P_{3/2}$ states, we observe similar deviations between our results and those from Refs. \cite{Safronova2016, Iskrenova2007} that we observed for the $8P_{1/2}$ and $8P_{3/2}$ states. It is important to note that in these calculations, mixture of many-body methods were employed to estimate the polarizabilities in which experimental energies were used to minimize the uncertainties. 

\section{Summary}

We have used linear response coupled-cluster theory in the relativistic framework to calculate the scalar and tensor polarizabilities for the ground and excited states of the Cs atom. Calculations are carried out at the singles- and doubles-excitation approximations and their uncertainties are estimated from the dominantly contributing triple contributions in the perturbative approach and QED effects. We also present results from random phase approximation and finite-order many-body perturbation theory to analyze propagation of the correlation effects in the evaluation of electric dipole polarizabilities. Our findings show that core polarization effects contribute dominantly over the pair-correlation contributions in the determination of electric dipole polarizabilities in Cs. Also, our finding shows that contributions from the Breit interactions are significant for precise estimations of the polarizabilities in the high-lying states in this atom. From the comparison between our results with the experimental data, we find good agreement among the relativistic coupled-cluster results with the measurements. Accuracy of the calculations are anticipated to be improved after inclusion of contributions from the full triple excitations in the relativistic coupled-cluster method.

\section*{Acknowledgment}

This work is supported by ANRF with grant no. CRG/2023/002558 and Department of Space, Government of India. All calculations reported in this work were performed on the ParamVikram-1000 HPC cluster at the Physical Research Laboratory (PRL), Ahmedabad, Gujarat, India.

\section*{Data availability}

The data that supports the findings of this study are available within the article.

\end{document}